\documentclass[pra,aps,twocolumn,superscriptaddress]{revtex4-1}

\usepackage[latin9]{inputenc}
\usepackage{graphicx,color,graphics}
\usepackage{dcolumn}
\usepackage{amsmath}
\usepackage{amssymb}
\usepackage{amsfonts}
\usepackage{hyperref}
\usepackage{url}
\usepackage{physics}

\newcommand\blankfootnote[1]{%
  \let\thefootnote\relax\footnotetext{#1}%
  \let\thefootnote\svthefootnote%
}

\begin{document}
\title{Spectroscopy of the $\mathbf{^1S_0}$-to-$\mathbf{^1D_2}$ clock transition in $^{176}$Lu$^+$}
\author{R. Kaewuam}
\affiliation{Centre for Quantum Technologies, National University of Singapore, 3 Science Drive 2, 117543 Singapore}
\author{T. R. Tan}
\affiliation{Centre for Quantum Technologies, National University of Singapore, 3 Science Drive 2, 117543 Singapore}
\affiliation{Department of Physics, National University of Singapore, 2 Science Drive 3, 117551 Singapore}
\author{K. J. Arnold}
\affiliation{Centre for Quantum Technologies, National University of Singapore, 3 Science Drive 2, 117543 Singapore}
\author{M. D. Barrett}
\email{phybmd@nus.edu.sg}
\affiliation{Centre for Quantum Technologies, National University of Singapore, 3 Science Drive 2, 117543 Singapore}
\affiliation{Department of Physics, National University of Singapore, 2 Science Drive 3, 117551 Singapore}
\date{\today}
\begin{abstract}
High precision spectroscopy of the $^1S_0$-to-${^1}D_2$ clock transition of $^{176}$Lu is reported.   Measurements are performed with Hertz level precision with the accuracy of the hyperfine-averaged frequency limited by the calibration of an active hydrogen maser to the SI definition of the second via a GPS link.  The measurements also provide accurate determination of the $^1D_2$ hyperfine structure. Hyperfine structure constants associated with the magnetic octupole and electric hexadecapole moments of the nucleus are considered, which includes a derivation of correction terms from third-order perturbation theory.
\end{abstract}

\pacs{Valid PACS appear here}
\maketitle


\section{\label{sec:intro}Introduction}
Singly-ionized lutetium is a unique atomic clock candidate supporting three clock transitions: a highly forbidden magnetic dipole (M1) transition $^1S_0$-to-${}^3D_1$ at 848 nm, a spin-forbidden electric quadrupole (E2) transition $^1S_0 $-to-$ {}^3D_2$ at 804 nm, and an E2 transition $^1S_0 $-to-$ {}^1D_2$ at 577 nm.  For each transition, hyperfine averaging eliminates shifts associated with the electronic angular momentum giving effective $J=0$ levels with low sensitivity to electro-magnetic fields \cite{Barrett2015, gan2018oscillating}.  Each transition has a unique sensitivity to environmental conditions such that frequency comparisons within the same apparatus provide important consistency checks for estimated systematic shifts.  

Transitions at 848 and 804\,nm have been observed \cite{Arnold2016,Kaewuam2017} and investigated \cite{Arnold2018}, which demonstrated competitive properties with leading clock candidates.  The 848-nm transition, in particular, offers an exceptionally low blackbody radiation (BBR) shift and all atomic properties relevant to clock performance offer an improvement over the Yb$^+$ octupole transition \cite{Arnold2018,gan2018oscillating}. Spectroscopy of the 577-nm transition has not yet been reported in the literature, but theoretical calculations \cite{porsev2018clock} have recently been carried out indicating a BBR shift competitive with the quadrupole transitions in Sr$^+$, Ca$^+$, Hg$^+$, and Yb$^+$.  Moreover, the calculated quadrupole moment of just $0.022\,e a_0^2$ could be managed without the need for averaging.

In addition to clock applications, measuring the hyperfine structure of the long-lived $^1D_2$ level offers the possibility of extracting the relatively unexplored magnetic octupole and electric hexadecapole moments of the nucleus as for $^3P_2$ levels discussed in \cite{BeloyA2008}.  In that work only leading second-order corrections arising from the coupling to a neighbouring $^3P_1$ level were considered.  Naively one might expect coupling to a singlet level to be diminished and hence the correction terms for the higher order nuclear moments minimal.  

In this paper we report high-resolution measurements of the $^1S_0$-to-${}^1D_2$ optical transitions in $^{176}$Lu$^+$ from which we extract hyperfine splittings with Hertz level accuracy.  Hyperfine structure constants associated with the nuclear magnetic octupole and electric hexadecapole moments are considered, which includes a derivation of correction terms up to third-order perturbation theory. Based on considerations for both $^1D_2$ and $^3D_2$, it is argued that evidence of higher order multipole moments should include consideration of leading order corrections from this third-order extension.  To our knowledge, such corrections have never been considered.  The feasibility of conclusively observing the influence of the nuclear octupole and electric hexadecapole moments in $^{176}$Lu$^+$ is also discussed.


The paper is organized as follows. A description of the experimental system is given in section~\ref{sec:experiment}, followed by the measurement procedures and results for $^1D_2$ in section~\ref{sec:measurements}. Then, a brief summary of relevant  hyperfine theory is given in section~\ref{sec:formalism} followed by its application to the $^1D_2$ and $^3D_2$ hyperfine structure in section~\ref{sec:hyperfineconstant}. 

\section{Experimental Setup}
\label{sec:experiment}
Measurements are performed in a four-rod linear Paul trap with axial end-caps as described in \cite{Kaewuam2017}.  Radial confinement is provided by a  $16.8\,\mathrm{MHz}$ radio-frequency (rf) potential applied to a pair of diagonally opposing rods via a quarter-wave helical resonator, a small dc voltage applied to the other pair of rods ensures a splitting of the transverse frequencies, and the end caps are held at $8\,\mathrm{V}$ to provide axial confinement. In this configuration, the secular trapping frequencies are $(\omega_x,\omega_y,\omega_z) = 2\pi\times (610,560,130)$ kHz, with $x,y$ indices denoting the two radial directions and $z$ the trap axis.  A magnetic field of $\sim 0.24\,\mathrm{mT}$ defines a quantization axis.

\begin{figure}
\includegraphics[width=0.95\linewidth]{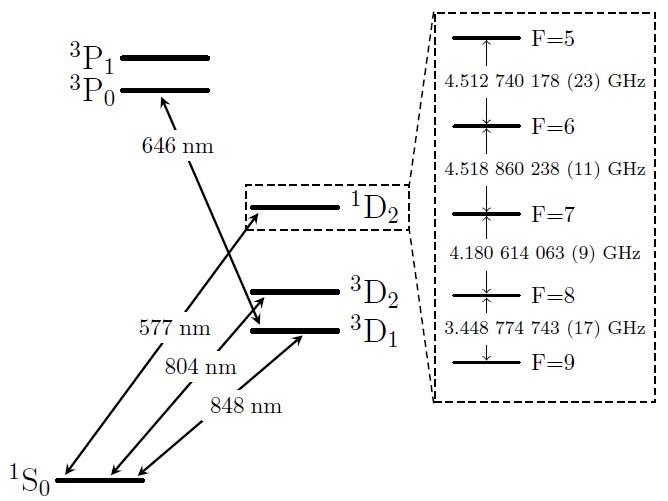}
\caption{Relevant energy level and transition diagram of a $^{176}$Lu$^+$ ion.  The hyperfine interaction gives rise to five hyperfine levels in $^1D_2$. The hyperfine splitting shown are determined from the measured transition frequencies of $^1S_0$-to-$^1D_2$. The uncertainties are the quadratic sum of statistical and systematic uncertainties.}
\label{LuTransition}
\end{figure}

The energy level structure of $^{176}$Lu$^+$ relevant to this work is shown in Fig.~\ref{LuTransition}.  There are three narrow linewidth optical transitions from the $^1S_0$ ground state to the upper $^3D_1$, $^3D_2$, and $^1D_2$ clock states.  The lifetime of $^3D_1$ is estimated to be approximately 172 hours \cite{Arnold2016}, and the lifetimes of $^3D_2$ and $^1D_2$ have been measured to be 17.3\,s and 180\,ms respectively \cite{Paez2016}. Doppler cooling and detection are achieved via scattering on the nearly closed $^3D_1 $-to-$ {}^3P_0$ transition at 646\,nm, which has a measured linewidth of $\sim 2\pi \times 2.45\,\mathrm{MHz}$ \cite{Kaewuam2017, Arnold2018Pol}.  Optical pumping into $^3D_1$ is facilitated by driving the $^1S_0 $-to-$ {}^3P_1$, $^3D_2 $-to-$ ^3P_1$, and $^1D_2 $-to-$ {}^3P_1$ transitions at 350 nm, 622 nm, and 895 nm, respectively.

\begin{figure}
\includegraphics[width=1\linewidth]{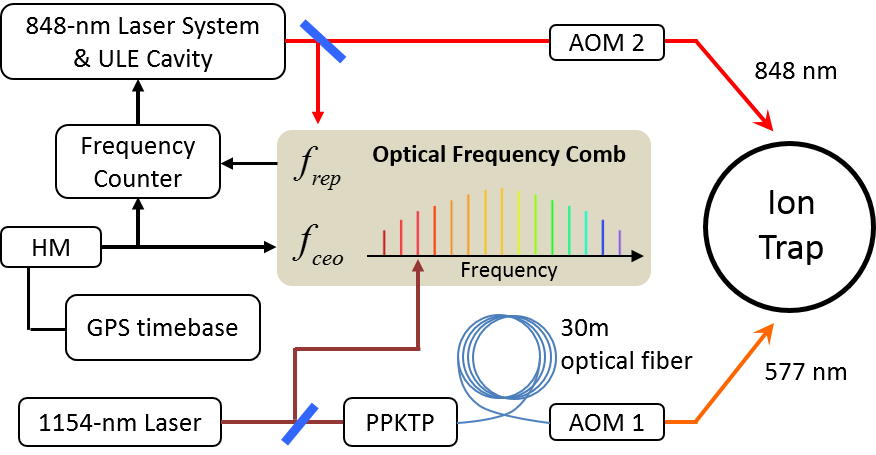}
\caption{Schematic diagram of the experiment. The optical frequency comb (OFC) is referenced to the 848-nm laser to stabilize $f_\mathrm{rep}$ and $f_\mathrm{ceo}$ is referenced to a HM. The 1154-nm laser is stabilized via phase locking to the OFC.}
\label{lasersetup}
\end{figure}

Spectroscopy of the $^1D_2$ level is implemented using a frequency-doubled extended-cavity-diode-laser system (ECDL) with a fundamental wavelength of $1154\,\mathrm{nm}$.  Second harmonic generation is accomplished using a fiberized, periodically-poled potassium titanyl phosphate (PPKTP) waveguide. The fundamental frequency is phase locked to an optical frequency comb (OFC). The short term ($<10\,$s) stability of the OFC is derived from a $\sim 1\,\mathrm{Hz}$ linewidth laser at 848 nm which is referenced to a 10 cm long ultra-low expansion (ULE) cavity with finesse of $\sim4\times10^5$. For longer times ($\gtrsim 10\,\mathrm{s}$), the OFC is steered to an active hydrogen maser (HM) reference. The frequency of the maser is calibrated to the SI (International System of Units) second by continuous comparison to a GPS timebase. The HM-GPS link exhibits a fractional instability of $2.8\times10^{-14}$/day. The 577-nm clock laser is switched with a double-passed acousto-optic modulator (AOM1 in in Fig. \ref{lasersetup}) which also controls the laser frequency relative to the comb.  A schematic of the experimental setup is shown in Fig. \ref{lasersetup}.

\section{Measurements}
\label{sec:measurements}
Spectroscopy of the $^1D_2$ level follows a similar procedure to that reported in \cite{Kaewuam2017}. The atom is first optically pumped to $^3D_1$ with success checked in real time using a Bayesian detection scheme reported in \cite{Arnold2016}.  When the atom is detected bright, the experiment continues with a $5\,\mathrm{ms}$ Doppler cooling pulse followed by optical pumping to $\ket{{^3}D_1, F=7,m=0}$.  A $\pi$-pulse on the 848\,nm clock transition is then applied to transfer the ion to $\ket{{^1}S_0, 7,\pm1}$.  The transfer efficiency of $\sim 95\%$ is limited by both state preparation of $\ket{{^3}D_1, 7,0}$ and the clock $\pi$-pulse. The fidelity of population transfer is improved to better than 99.9\% by state detection after the $\pi$-pulse. If the ion is detected bright, state preparation of $\ket{{^3}D_1, 7,0}$ and shelving to $\ket{{}^1S_0,7,\pm 1}$ is repeated. When a dark state is confirmed, the experiment proceeds with a $\pi$-polarized 577-nm clock pulse to drive a transition to the $\ket{{^1}D_2, F^{\prime},0}$ state, where $F^{\prime} = 5, 6,...,9$. Population remaining in $^1S_0$ is reshelved to $^3D_1$ with approximately 98\% fidelity via a $\pi$-pulse from the 848-nm laser and subsequently detected by 646-nm fluorescence.

\begin{figure}
\includegraphics[width=1\linewidth]{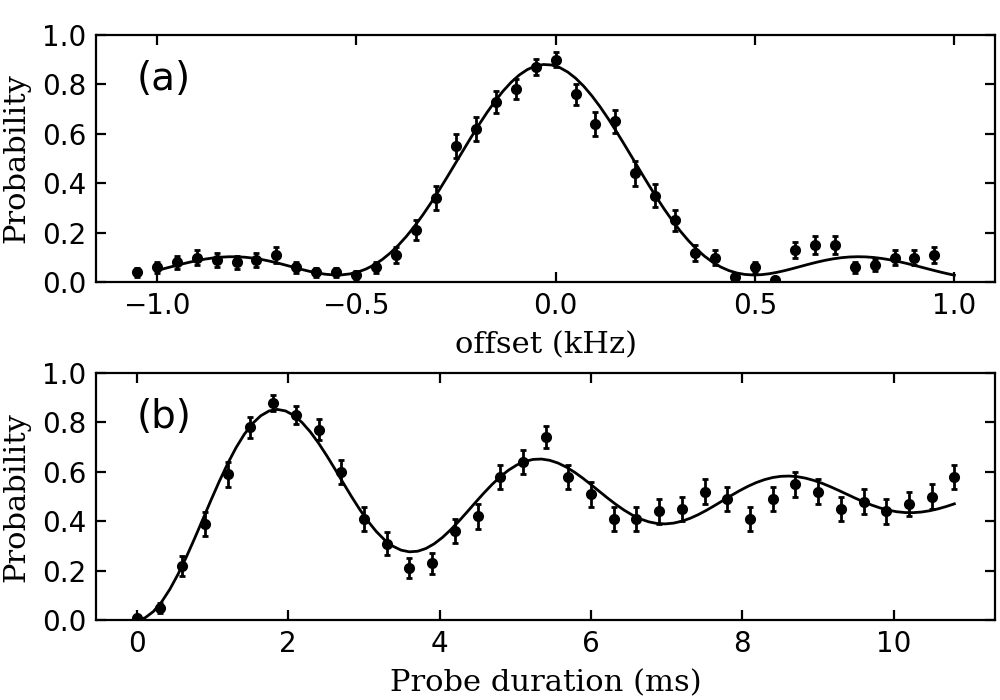}
\caption{Rabi spectroscopy of the transition from $\ket{{}^1S_0 ,7, +1}$ to $\ket{{}^1D_2,6,0}$. Each point represents an average of 100 experiments. (a) Transition probability as a function of laser frequency offset for a fixed pulse time of 2\,ms. (b) Rabi oscillation on the carrier transition. The model curve assumes dephasing of the oscillation due to thermal motion and an overall exponential damping, which accounts for the unstabilized optical fiber path.}
\label{Interragation}
\end{figure}

The transition probability when driving $\ket{{^1}S_0,7,+1}$ to $\ket{{^1}D_2, 6,0}$ as a function of either laser frequency offset or probe time is shown in Fig.~\ref{Interragation}. The limited coherence time indicated in Fig.~\ref{Interragation}(b) is likely limited by both the unstabilized $\approx$ 30 m optical fiber path from the laser source to the location of the ion and thermal dephasing. Nevertheless the 2\,ms interrogation time provides sufficient frequency resolution to resolve the ground state splitting. 

To remove the first-order Zeeman shift of the ground state, the frequency of the 577-nm laser is steered to the average of a pair of Zeeman transitions from $\ket{{^1}S_0,7,\pm1}$ to $\ket{{^1}D_2, F',0}$, using a servo technique similar to \cite{Bernard1998} that alternatively interrogates either side of each transition to derive error signals. There are four interrogations in total: two for $m_F = +1$ and two for $m_F = -1$. From these four points the laser frequency offset from the average transition frequency and splitting of two Zeeman components can be extracted. The appropriate error correction is applied to the oscillator driving AOM1 and recorded by a computer program. The servo to each of the hyperfine lines is implemented for a duration of $\approx$ 1000\,s, and the corresponding statistical uncertainty of a measured optical frequency is at the $\sim$1\,Hz level limited by the flicker noise floor of the HM.  The frequency offset and linear frequency drift of the HM were assessed by comparison with a GPS-linked reference over a 6 month period, yielding a fractional uncertainty of $3\times10^{-15}$ of the HM frequency.

Interleaved with the 577-nm servo loop is a similar servo loop that steers the 848-nm laser to the average of transitions from $\ket{{^3}D_1, 7,0}$ to $\ket{{^1}S_0,7,\pm1}$. This is serves two purposes, (i) to determine the amplitude of the magnetic field from the difference of the two Zeeman transitions and (ii) provide a consistently check of the absolute frequency measurement methodology used for the 577-nm measurements. At each different 577-nm line measurement, which were taken over the course three days, the frequency of this 848-nm reference transition is measured from the interleaved servo. The systematic shifts affecting the 848-nm transition are expected to have instability well below the $1\times10^{-15}$ level. The five reference measurements have an rms spread of 0.6\,Hz, or $1.7\times10^{-15}$ fractionally. This is consistent with statistical uncertainty of each measurement which is limited to $2\times10^{-15}$ by the stability of the HM for a averaging time of $10^{3}$ seconds.  Additionally, a measurement of the 848-nm reference transition was repeated four months later and found to be in agreement at $2\times10^{-15}$ (see Supplemental Material).  This supports the $3\times10^{-15}$ uncertainty assessment of the HM frequency from the GPS-link. 

For the $^1S_0$-to-$^1D_2$ transitions, systematic shifts at the Hz level are completely determined by quadratic Zeeman shifts:  probe-induced ac-Stark shifts are negligibly small owing to the relatively short lifetime, the quadrupole moment for the $^1D_2$ is approximately two-orders of magnitude smaller than for the triplet states, and micromotion is trivially controlled to the sub-Hz level.  Expressions for the quadratic Zeeman shift are identical to those for the $^3D_2$ given in \cite{Kaewuam2017} differing only in the value of $g_J$.  For convenience these expressions are given in Appendix~\ref{QuadZeeman}.

The Zeeman splitting between $\ket{{}^1S_0,7,\pm 1}$ transitions is inferred from the 848-nm servo, to assess the strength of the magnetic field, $B$. As the laser couples both $m_F = \pm 1$ ground states to the upper $m_F'=0$ state, the transition frequency of one Zeeman transition is shifted due to off-resonant coupling to the other Zeeman transition. The shifts are equal and opposite for the two transitions and thus do not lead to an overall shift in the Zeeman-averaged transition frequency.  However, the accuracy in determining $B$ can be degraded. With the interrogation time of $\approx$ 5\,ms, the effect of off-resonant coupling is at the $0.1\%$ level.  This is much smaller than the accuracy of $g_I = -2.436 \times 10^{-4}$ \cite{Ritter1962,Brenner1985} of the $^1S_0$ level, which is assumed to be 0.5\%.  From the measured Zeeman splitting an average field of 0.2386(12)\,mT is deduced.

Within the LS-coupling limit, $g_J$ for $^1D_2$ would be given by $g_L=1$.  Recent calculations have given $g_J=1.01$ \cite{porsev2018clock} indicating that mixing does not significantly influence the value.  Consequently, we take the calculated value for $g_J$ in determining the Zeeman shifts and assume the error is dominated by the magnetic field determined from the ground state.  The resulting Zeeman-corrected optical frequencies, $\nu_{F^{\prime}}$, for each of the transitions from $\ket{{^1}S_0,7,0}$ to $\ket{{^1}D_2,F^{\prime},0}$ are, in Hz units: 
\begin{subequations}
\begin{eqnarray}
\nu_5 &=& 519\,622\,296\,515\,663.7\, \pm(1.6)_\mathrm{stat}\pm(31.2)_\mathrm{z},\\
\nu_6 &=& 519\,617\,783\,775\,485.6\,\pm(1.6)_\mathrm{stat}\pm(8.5)_\mathrm{z},\\
\nu_7 &=& 519\,613\,264\,915\,247.9\,\pm(1.6)_\mathrm{stat}\pm(2.1)_\mathrm{z},\\
\nu_8 &=& 519\,609\,084\,301\,184.6\,\pm(1.6)_\mathrm{stat}\pm(10.4)_\mathrm{z},\\
\nu_9 &=& 519\,605\,635\,526\,441.6\,\pm(1.6)_\mathrm{stat}\pm(27.3)_\mathrm{z},
\end{eqnarray}
\label{eq:ResultFreq}
\end{subequations}
where the values in $(...)_\mathrm{stat}$ denote statistical uncertainties in optical frequency measurement while $(...)_\mathrm{z}$ are systematic uncertainties from the quadratic Zeeman shifts. As the systematic shifts arise from imperfect knowledge of the magnetic-field, they are strongly correlated.  In particular the average frequency has a calculated magnetic field sensitivity of $1.2\times10^{-16}/\mathrm{mT}^2$.  Consequently, the average frequency here is limited completely by the statistical error.


\section{\label{sec:formalism}Hyperfine interaction theory}
The accuracy of the measurements made allow high accuracy determination of hyperfine splittings, which are suitable for investigating hyperfine structure constants.  For this purpose, we give a summary of relevant theory.  We follow closely the work of Woodgate \cite{Woodgate1966} and Beloy \cite{BeloyA2008}, and include an extension to third order correction terms.

A nucleus can be approximately described as a point-like collection of electromagnetic moments. From the relativistic treatment in \cite{BeloyA2008}, the hyperfine Hamiltonian can be written as a sum of multipole interactions between electrons and nucleons,
\begin{equation}
H_{\mathrm{hfs}} = \sum_{k=1}^\infty\vb*{T}^e_k\cdot\vb*{T}^n_k = \sum_{k=1}^\infty\sum_{\mu=-k}^k(-1)^\mu \vb*{T}^e_{k,\mu}\vb*{T}^n_{k,-\mu},
\end{equation}
where  $\vb*{T}^e_{k,\mu}$ and $\vb*{T}^n_{k,\mu}$ are spherical tensor operators of rank $k$ that operate on the space of electronic and nuclear coordinates, respectively. The sum excludes the term $k=0$ because the monopole interaction is included in the unperturbed atomic Hamiltonian. Basis states where the total angular momentum $\vb*{F} = \vb*{I}+\vb*{J}$ is conserved are denoted $\ket{\gamma IJFm_F}$ where $\gamma$ denotes all other quantum numbers. From to the Wigner-Eckart theorem, a matrix element of $H_{\mathrm{hfs}}$ over the basis set is,
\begin{multline}
\matrixel{\gamma' IJ'F'm_F'}{H_{\mathrm{hfs}}}{\gamma IJFm_F}
= \delta_{FF'}\delta_{m_F'm_F}(-1)^{J'+I+F}\\
\times\sum\limits_{k=1}^{k'} \left\{\begin{matrix}
F & J' & I \\
k & I & J
\end{matrix}\right\}\matrixel{\gamma' J'}{|\vb*{T}^e_{k}|}{\gamma J}\matrixel{I}{|\vb*{T}^n_{k}|}{I},
\label{eq:melHFI}
\end{multline}
where $k'=\boldsymbol{\mathrm{min}}(2I,J+J')$. Following the notations used in \cite{Woodgate1966}, the energy shift of a level with specific quantum numbers $\gamma$, $J$, and $F$ can be expressed, 
\begin{multline}
\label{eq:hyperfineShift}
W_{JF} = \sum_{k=1}^{k_\mathrm{m}} X_k(IJF)U_k^{(1)}(J)\\
+\sum_{k=0}^{k_\mathrm{m}}\left( X_k(IJF)\sum_{n=2}^\infty U_k^{(n)}(J)\right),
\end{multline}
where the $F$-dependent scale factor is given by,
\begin{equation}
\label{hyperfineShift1}
 X_k(IJF) = (-1)^{I+J+F}\frac{\left\{\begin{matrix}
F & J & I \\
k & I & J
\end{matrix}\right\}}{ \left(\begin{matrix}
I & k & I \\
-I & 0 & I
\end{matrix}\right)
 \left(\begin{matrix}
J & k & J \\
-J & 0 & J
\end{matrix}\right)},
\end{equation}
$k_\mathrm{m}=\boldsymbol{\mathrm{min}}(2I,2J)$ is the minimum number of electromagnetic poles of either the relevant electronic state or the nucleus, and $U_k^{(n)}$ are terms arising from $n^\mathrm{th}$-order perturbation theory.  The form of Eq.~\ref{eq:hyperfineShift} up to second-order was first derived by Woodgate \cite{Woodgate1966}.  Following that work, an outline of the derivation for $U_k^{(3)}(J)$ and the extension to all orders is given in Appendix~\ref{hyperfine} along with explicit expressions for $U_k^{(n)}$ up to $n=3$. 

Similar to the first-order correction, $X_k(IJF)$ appears as an overall multiplication factor for other perturbative corrections ($n>1$), except the index $k$ starts from $k=0$ in the summation. Since $X_0(IJF) = 1$ for all possible quantum numbers $F$, $J$, and $I$, the energy shift is,
\begin{equation}
W_{JF} = \sum_{n=2}^\infty U_0^{(n)}(J)+ \sum_{k=1}^{k_\mathrm{m}}\sum_{n=1}^\infty X_k(IJF) U_k^{(n)}(J).
\label{eq:HFIshift}
\end{equation}
The first term in Eq.~(\ref{eq:HFIshift}) implies an overall shift, which should properly be considered an isotope shift similar to the shifts arising from the finite size and finite mass of the nucleus \cite{King1984}. To second order, this overall shift has been pointed out and discussed in \cite{BeloyB2008}.  The $F$-dependent factor $X_k(IJF)$ entering in the first- and second-order corrections in an identical way was first noted by Woodgate \cite{Woodgate1966} and much later by Beloy and Derevianko \cite{BeloyB2008}.  Woodgate interpreted $U_k^{(2)}(J)$ as a second-order correction to $U_k^{(1)}(J)$, which has a direct relation to the conventional hyperfine constants $A, B, C,...$ defined by,
\begin{subequations}
\label{hyperfineConstants}
\begin{eqnarray}
A &=& \frac{1}{IJ} U_1^{(1)}(J),\\
B &=& 4U_2^{(1)}(J),\\
C &=& U_3^{(1)}(J),\\
D &=& U_4^{(1)}(J).
\end{eqnarray}
\end{subequations}
Notationally, it is convenient to introduce 
\begin{equation}
\label{hyperfineTrue}
U_k(J) = \sum_{n=1}^\infty U_k^{(n)}(J),
\end{equation}
with $U_0^{(1)}(J)\equiv0$.  Equation~\ref{eq:HFIshift} then has the simple form
\begin{equation}
W_{JF} =\sum_{k=0}^{k_\mathrm{m}} X_k(IJF) U_k(J).
\label{eq:HFIshift2}
\end{equation}
with the $k=0$ term being a scalar, hyperfine-induced isotope shift.  Hyperfine constants $A',B',C',...$ related to $U_k(J)$ by equations analogous to Eqs.~\ref{hyperfineConstants} can then be determined exactly from the hyperfine splittings.  Throughout the literature $A', B',...$ would be referred to as the uncorrected hyperfine structure constants with corrections made to accommodate the definitions given by Eqs~\ref{hyperfineConstants}.

\section{Hyperfine constants of $^1D_2$ and $^3D_2$}
\label{sec:hyperfineconstant}
Using the expressions given in Sec. \ref{sec:formalism}, the energy shift $W_F$ due to the hyperfine interaction for $D_2$ level in terms of (uncorrected) hyperfine constants can be readily determined. For both $^1D_2$ and $^3D_2$, equations for the hyperfine splittings $\delta W_F = W_F - W_{F-1}$ are given by
\begin{subequations}
\label{eq:HFS}
\begin{eqnarray}
\delta W_6 &=& 6A'-\tfrac{153}{364}B'+\tfrac{459}{91}C'-\tfrac{14535}{1001}D'\\
\delta W_7 &=& 7A'-\tfrac{25}{104}B'-\tfrac{21}{13}C'+\tfrac{285}{13}D'\\
\delta W_8 &=& 8A'+\tfrac{5}{91}B'-\tfrac{368}{91}C'-\tfrac{1520}{91}D'\\
\delta W_9 &=& 9A'+\tfrac{27}{56}B'+\tfrac{27}{7}C'+\tfrac{45}{7}D'
\end{eqnarray}
\end{subequations}
 with the inverse relationships
 \begin{eqnarray*}
 A' &=& \tfrac{11}{525}\delta W_6 + \tfrac{51}{1400} \delta W_7 + \tfrac{117}{2800} \delta W_8 + \tfrac{19}{600} \delta W_9,\\
 B' &=& -\tfrac{88}{105}\delta W_6 - \tfrac{5}{7}\delta W_7 + \tfrac{39}{238}\delta W_8 +\tfrac{247}{255}\delta W_9\\
 C' &=& \tfrac{11}{150}\delta W_6 - \tfrac{7}{200}\delta W_7 -\tfrac{299}{3400}\delta W_8 + \tfrac{1729}{30600} \delta W_9\\
 D' &=& -\tfrac{11}{1050}\delta W_6 + \tfrac{33}{1400}\delta W_7 - \tfrac{429}{23800}\delta W_8 + \tfrac{143}{30600}\delta W_9
 \end{eqnarray*}
From the measured optical frequencies given in Eq.~(\ref{eq:ResultFreq}), the uncorrected hyperfine constants for the $^1D_2$ level are, in Hz units,
\begin{subequations}
\begin{eqnarray}
A' &=& -543\, 069\, 419.3 \pm (1.7)_\mathrm{z}\pm(0.07)_\mathrm{stat}\\
B' &=& 2\, 984\, 226\, 871.4 \pm(8.8)_\mathrm{z}\pm(2.8)_\mathrm{stat}\\
C' &=& 6904.2 \pm(1.5)_\mathrm{z}\pm(0.3)_\mathrm{stat}\\
D' &=& -42.018\pm(58)_\mathrm{z}\pm(95)_\mathrm{stat}.
\end{eqnarray}
\end{subequations}
For comparison, using the optical frequencies measured in \cite{Kaewuam2017}, the corresponding uncorrected hyperfine constants for the $^3D_2$ level are, in Hz units,
\begin{subequations}
\begin{eqnarray}
A' &=& 1\, 370\, 376\, 728 (8)\\
B' &=& 1\, 825\, 831\,163 (350)\\
C' &=& 396\, 959(42)\\
D' &=& -1824(12).
\end{eqnarray}
\end{subequations}
The value of $D'$ for $^3D_2$ is slightly different than that reported in \cite{Kaewuam2017} owing to an incorrect $g_J$ factor used in the evaluation of Zeeman shifts.  Appropriately corrected optical frequencies are tabulated in Appendix~\ref{QuadZeeman}.

It should be noted that there is a large cancellation of systematic shifts in the determination of these coefficients.  The $D'$ coefficient in particular has a Zeeman dependence of just $-180\,\mathrm{Hz/mT^2}$ and $100\,\mathrm{Hz/mT^2}$ for $^3D_2$ and $^1D_2$ respectively.  Consequently the statistical errors can be significant even if the quadratic Zeeman shift systematic dominates the frequency uncertainty.  Statistical errors are much larger for the $^3D_2$ measurements owing to the stability of the optical comb used at that time.

To compare these values to the usual hyperfine constants proportional to the appropriate electromagnetic moment of the nucleus, correction terms must be calculated.  As noted in \cite{BeloyB2008}, leading order dipole-dipole (d-d), and dipole-quadrupole (d-q) corrections do not contribute to $D$ and only d-q corrections apply to $C$.  This has been proved as special cases in \cite{BeloyA2008, beloy2008nuclear}, but the general second-order correction derived by Woodgate \cite{Woodgate1966} and given in Eq.~\ref{eq:2ndOrder} makes this immediate from the 6-j symbols involved.

Including only d-q corrections, the corrected $C$ coefficient for the $^1D_2$ level is given by
\begin{equation}
\label{Ceq}
C = C' +\zeta({^3}D_1)-\sqrt{\tfrac{3}{7}}\zeta({^3}D_2)+\tfrac{1}{7}\sqrt{\tfrac{3}{2}}\zeta({^3}D_3),
\end{equation}
where
\begin{equation}
\zeta({^3}D_J) = \tfrac{6}{175}\mu Q\,\frac{\langle {^3}D_J \| \vb*{T}^e_{1} \| {^1}D_2\rangle \langle{^3}D_J\| \vb*{T}^e_{2}\| {^1}D_2\rangle}{E_{^1D_2}-E_{{^3}D_J}},
\end{equation}
with $\mu$ and $Q$ being the usual magnetic dipole and electric quadrupole moment of the nucleus respectively.  A similar expression can be found for the $^3D_2$ level by interchanging $^1D_2$ and $^3D_2$ in all expressions.

Matrix elements for these corrections are not available for all contributions. However, matrix elements for contributions from $^3D_1$ have been computed for the purposes of estimating hyperfine quenching rates of the $^3P_0$ level used for state detection \cite{Paez2016}.  Relevant values are tabulated in Table~\ref{MEs}.  In the case of the $^3D_2$ level these give a correction of approx 360(100)\,kHz, which has the same sign and magnitude of $C'$.  As other terms have a larger energy denominator and smaller coefficient in Eq.~\ref{Ceq}, we would not expect these to provide a large cancellation indicating a fairly large value for $C$.  For the $^1D_2$ level, the corresponding contribution from $^3D_1$ is $\sim -2.4\,\mathrm{kHz}$ with an error bar of $50\%$.  This has the opposite sign as $C'$ so leads to a some cancellation.  Thus it would appear there is likely a very large difference in the $C$ coefficient for the triplet and singlet $J=2$ levels.  
\begin{table}[ht]
\caption{\label{MEs} Matrix elements of the electronic operators $\vb*{T}^e_{1}$ and $\vb*{T}^e_{2}$ in units of MHz/$\mu_N$ and MHz/barn, respectively.  Values are taken from \cite{Paez2016}\footnote{Signs were not given \cite{Paez2016} and one matrix element was missing.  These were given in a private communication.}.  Note that signs of the matrix elements depend on a convention choice but relative signs between them are fixed by that choice.}
\begin{ruledtabular}
\begin{tabular}{lr}
\multicolumn{1}{c}{\hspace{1cm}ME}                                                              & \multicolumn{1}{c}{Value   \hspace{1.2cm}}  \\
\hline \\ [-0.3pc]
\hspace{1cm}$\left\langle {}^3D_1\|\,\vb*{T}^e_{1}\| {}^1D_2 \right\rangle$      & -10620\,(870) \hspace{1cm} \\[0.3pc]
\hspace{1cm}$\left\langle {}^3D_1 ||\,\vb*{T}^e_{2}|| {}^1D_2 \right\rangle$      &  70\,(45) \hspace{1cm} \\[0.3pc]

\hspace{1cm}$\left\langle {}^3D_1 ||\,\vb*{T}^e_{1}|| {}^3D_2 \right\rangle$      & 18680\,(1900) \hspace{1cm} \\[0.3pc]
\hspace{1cm}$\left\langle {}^3D_1 ||\,\vb*{T}^e_{2}|| {}^3D_2 \right\rangle$      & 700\,(100) \hspace{1cm} \\[0.3pc]

\hspace{1cm}$\left\langle {}^3D_1 ||\,\vb*{T}^e_{2}|| {}^3D_3 \right\rangle$      & 200\,(50) \hspace{1cm} \\[0.3pc]
\end{tabular}
\end{ruledtabular}
\end{table}

For the $D$ coefficient, leading second-order corrections are q-q and possibly dipole-octupole (d-o).  Including only these terms the expression for the $^1D_2$ level is given by
\begin{multline}
D=D'+\xi({^3}D_1)-\tfrac{3}{7}\xi({^3}D_2)+\tfrac{3}{28}\xi({^3}D_3)\\
-\chi({^3}D_1)+\tfrac{1}{\sqrt{6}}\chi({^3}D_2)-\tfrac{1}{3\sqrt{14}}\chi({^3}D_3),
\end{multline}
where
\begin{equation}
\xi({^3}D_J) = \tfrac{33}{15925} Q^2\,\frac{|\langle {^3}D_J \| \vb*{T}^e_{2} \| {^1}D_2\rangle|^2}{E_{^1D_2}-E_{{^3}D_J}},
\end{equation}
and
\begin{multline}
\chi({^3}D_J) = \tfrac{11}{245}\sqrt{\tfrac{2}{7}}\,\mu \Omega\\
\times\frac{\langle {^3}D_J \| \vb*{T}^e_{1} \| {^1}D_2\rangle \langle{^3}D_J\| \vb*{T}^e_{3}\| {^1}D_2\rangle}{E_{^1D_2}-E_{{^3}D_J}}.
\end{multline}
with $\Omega$ the magnetic octupole moment of the nucleus as defined in \cite{BeloyA2008}.  Expressions for $^3D_2$ can again be obtained by interchanging $^1D_2$ and $^3D_2$.

Only $\xi({}^3D_1)$ can be estimated from the given matrix elements giving $\sim 1280(430)\,\mathrm{Hz}$ and $\sim 1.5\,\mathrm{Hz}$ for the $^3D_2$ and $^1D_2$ levels, respectively.  The correction for $^1D_2$ is only accurate to about a factor of 3 but is clearly much smaller than $D'$ in this case.  Since $\xi({}^3D_J)$ is positive definite there will be some cancellation of other q-q corrections, but it is unclear if d-o corrections would contribute significantly.  

Leading third order corrections for the $D$ coefficient would be d-d-q corrections and these should not be disregarded.  As a crude estimate, such terms would have the scale of a d-q correction term for $C$ multiplied by the ratio of a $\vb*{T}^e_{1}$ matrix element and an energy separation.  Including the magnetic dipole moment, the matrix element is on the order 10\,GHz and energy separations are on the order of 10\,THz.  Hence, we might expect d-d-q corrections to be on the order of $10^{-3}$ of the d-q corrections for $C$.  For both $^1D_2$ and $^3D_2$ these would be of similar magnitude to the estimated q-q corrections for the respective $D$ coefficient.  For the same reasons, d-d-d corrections to $C$ may also be important.  To our knowledge, third order corrections have never been considered.

Even if all required matrix elements were calculated, evidence of the higher order multipole moments would be contingent on the validity of those calculations and, ideally, that validity should be experimentally tested.  Since the matrix elements essentially determine a hyperfine mixing between fine-structure levels, any measurable consequence of that mixing could serve as a test of the theory.  For Lu$^+$, such mixing would give rise to: (a) decays from $^3P_0$ to levels other than $^3D_1$, (b) deviations of the $g$-factors applicable to the upper state and (c) forbidden transitions from the ground-state such as E2 transitions to $^3D_1$  or M1 transitions to $^1D_2$ or $^3D_2$.  Measured rates for quenching decays from $^3P_0$ in $^{175}$Lu$^+$ were reported in \cite{Paez2016}.  Quadrupole transitions from $^3D_1$ to $^1S_0$ have also been observed and coupling strengths could be readily calibrated.  Ultimately it is desirable to have high precision measurements of $g$-factors for the assessment of magnetic fields and the average of $g_F$ over all hyperfine levels is $g_I$ \cite{Barrett2015}.  Thus, there is opportunity to rigorously test the accuracy of correction terms.

\section{Summary}
In this paper we have performed high resolution spectroscopy of the $^1S_0$-to-$^1D_2$ clock transition in $^{176}$Lu$^+$.  Transitions to all hyperfine levels have been measured to Hertz level precision. This sets the stage for clock operation incorporating hyperfine averaging in which the laser is servoed over all five transitions measured here.  Limited knowledge of $g_I$ limits the current accuracy of individual transitions but this uncertainty can be significantly reduced with improved assessment of $g_I$.  Moreover, hyperfine averaging practically eliminates the systematic uncertainty associated with the static magnetic field.  As a by-product of this work we have extracted accurate determinations of the hyperfine structure.

Having accurate assessments of the hyperfine splittings for $^1D_2$ and $^3D_2$ prompted us to investigate the possible influence of higher order nuclear moments, specifically the magnetic octupole and electric hexadecapole moment.  To this end we have extended previous theory work to include third order perturbation theory.  We have argued that proper analysis of the higher-order nuclear moments should consider at least the leading order terms that appear in the third-order result.  

In the case of $^{176}$Lu$^+$, it is unclear if theory could attain sufficient accuracy to allow conclusive confirmation on the existence of the higher order nuclear moments, but experiments have been suggested that could at least test the validity of theoretical results.  Similar such experiments would be applicable in any system claiming to have observed these properties.  In the case of $^{176}$Lu$^+$, the experimental tests would be readily accessible as the system is developed towards a high performance clock.
 
\acknowledgements
We thank Vladimir Dzuba, Marianna Safronova and Sergey Porsev for useful discussions.  This work is supported by the National Research Foundation, Prime Ministers Office, Singapore and the Ministry of Education, Singapore under the Research Centres of Excellence programme. This work is also supported by A*STAR SERC 2015 Public Sector Research Funding (PSF) Grant (SERC Project No: 1521200080). T. R. Tan acknowledges support from the Lee Kuan Yew post-
doctoral fellowship.
\appendix
\section{Quadratic Zeeman shifts.}
\label{QuadZeeman}
Expressions for the quadratic Zeeman shifts for both the $^1D_2$ and $^3D_2$ are functionally equivalent differing only in the value of $g_J$ and the hyperfine splittings.  Defining the hyperfine interval $\delta \nu_{F'} = \nu_{F'} - \nu_{F'-1}$ in units of frequency, the quadratic shifts $\Delta_{F'}$ for the $m_F' = 0$ states are:
\begin{subequations}
\begin{eqnarray}
\Delta E_5/h&=& -\frac{16}{13}\frac{\mathcal{R}}{\delta \nu_6},\\
\Delta E_6/h&=& \frac{16}{13}\frac{\mathcal{R}}{\delta \nu_6} - \frac{102}{65}\frac{\mathcal{R}}{\delta \nu_7},\\
\Delta E_7/h&=& \frac{102}{65}\frac{\mathcal{R}}{\delta \nu_7} -\frac{117}{85}\frac{\mathcal{R}}{\delta \nu_8},\\
\Delta E_8/h&=& \frac{117}{85}\frac{\mathcal{R}}{\delta \nu_8} -\frac{14}{17}\frac{\mathcal{R}}{\delta \nu_9},\\
\Delta E_9/h&=& \frac{14}{17}\frac{\mathcal{R}}{\delta \nu_9}.
\end{eqnarray}
\end{subequations}
where $\mathcal{R} = (g_J-g_I)^2\mu_B^2 B^2/h^2$ with $g_J$ and $g_I$ the usual $g$-factors. Within the LS-coupling limit $g_J=1$ and $7/6$ for $^1D_2$ and $^3D_2$, respectively.

In \cite{Kaewuam2017}, the calculated Zeeman shifts inadvertently used a value of $g_J=1/2$.  The appropriately corrected optical frequencies are, in Hz,
\begin{subequations}
\begin{eqnarray}
\nu_5 &=& 372\,776\,905\,829\,552\,(200),\\
\nu_6 &=& 372\,784\,362\,667\,641\,(200),\\
\nu_7 &=& 372\,793\,515\,721\,790\,(200),\\
\nu_8 &=& 372\,804\,577\,481\,195\,(200),\\
\nu_9 &=& 372\,817\,792\,702\,607\,(200).
\end{eqnarray}
\label{eq:ResultFreq804}
\end{subequations}
\section{Third-order hyperfine corrections}
\label{hyperfine}
In this section an outline of the third-order correction to the HFS is given illustrating that it has the same $F$-dependent factor $X_k(IJF)$ as the first- and second-order terms.  The derivation illustrates how the form of the perturbation can be extended to all orders of perturbation theory.  For completeness, expressions for the first- and second-order terms are also given, which also establishes notation. 

Using the notation $\mathcal{I}^{(k)} \equiv \langle I \| \vb*{T}^n_{k}\| I\rangle$ and $\mathcal{Q}^{(k)}_{J_1J_2} \equiv \langle J_1\| \vb*{T}^e_{k}\| J_2\rangle$ for the reduced nuclear and electronic matrix elements, respectively, expressions for $U_k^{(1)}(J)$ and $U_k^{(2)}$ are given by \cite{Woodgate1966},
\begin{equation}
U_k^{(1)}(J) = 
\left(\begin{matrix}
I & k & I \\
-I & 0 & I
\end{matrix}\right)\left(\begin{matrix}
J & k & J \\
-J & 0 & J
\end{matrix}\right)\mathcal{I}^{(k)}\mathcal{Q}^{(k)}_{JJ},
\end{equation}
and
\begin{multline}
\label{eq:2ndOrder}
U_k^{(2)}(J)=
\left(\begin{matrix}
I & k & I \\
-I & 0 & I
\end{matrix}\right)\left(\begin{matrix}
J & k & J \\
-J & 0 & J
\end{matrix}\right)\\
\times \sum_{k_1k_2}\Bigg[(-1)^{2I+2J+k_1+k_2+k}(2k+1)\\
\times
 \left\{\begin{matrix}
k_1 & k_2 & k \\
I & I & I
\end{matrix}\right\}\mathcal{I}^{(k_1)}\mathcal{I}^{(k_2)}\\
\times \sum_{J'}
 \left\{\begin{matrix}
k_1 & k_2 & k \\
J & J & J'
\end{matrix}\right\}\frac{\mathcal{Q}^{(k_1)}_{JJ'}\mathcal{Q}^{(k_2)}_{J'J}}{(E_J-E_{J'})}\Bigg],
\end{multline}
respectively, where $E_J$ is the unperturbed fine-struture energy.  

Using Eq. (\ref{eq:melHFI}), the third-order correction to the energy $W_{JF}$ can be written in the compact form
\begin{equation}
W^{(3)}_{JF} = \sum_{J'J''}\frac{\mathcal{P}-\mathcal{G}}{(E_J-E_{J'})(E_J-E_{J''})},
\end{equation}
where $\mathcal{P}$ and $\mathcal{G}$ are given by 
\begin{widetext}
\begin{subequations}
\label{eq:PG}
\begin{align}
\label{eq:P}
\mathcal{P} &= \sum_{k_1k_2k_3}(-1)^{3I+J+J'+J''+3F}
\left\{\begin{matrix}
F & J & I \\
k_1 & I & J'
\end{matrix}\right\}
\left\{\begin{matrix}
F & J' & I \\
k_2 & I & J''
\end{matrix}\right\}
\left\{\begin{matrix}
F & J'' & I \\
k_3 & I & J
\end{matrix}\right\}\mathcal{I}^{(k_1)}\mathcal{I}^{(k_2)}\mathcal{I}^{(k_3)}\mathcal{Q}^{(k_1)}_{JJ'}\mathcal{Q}^{(k_2)}_{J'J''}\mathcal{Q}^{(k_3)}_{J''J}\\
\label{eq:G}
\mathcal{G} &= \delta_{J'J''}\sum_{k_1k_2k_3}(-1)^{3I+2J+J'+3F}
\left\{\begin{matrix}
F & J & I \\
k_1 & I & J
\end{matrix}\right\}
\left\{\begin{matrix}
F & J & I \\
k_2 & I & J'
\end{matrix}\right\}
\left\{\begin{matrix}
F & J' & I \\
k_3 & I & J
\end{matrix}\right\}\mathcal{I}^{(k_1)}\mathcal{I}^{(k_2)}\mathcal{I}^{(k_3)}\mathcal{Q}^{(k_1)}_{JJ}\mathcal{Q}^{(k_2)}_{JJ'}\mathcal{Q}^{(k_3)}_{J'J}.
\end{align}
\end{subequations}
Following Woodgate \cite{Woodgate1966}, the \textit{Biedenharn-Elliott} identity \cite[pg 305, Eq.~{7}]{varshalovich1988quantum} can be used to reduce the number of 6j-symbols having an $F$ dependence.  Explicitly, the last two 6j-symbols in Eq.~\ref{eq:P} can be written
\begin{equation*}
(-1)^{J+J'+J''+3I+F} \begin{Bmatrix} J' & I & F\\ I & J'' & k_2 \end{Bmatrix} \begin{Bmatrix} J & I  & F\\ I & J'' & k_3 \end{Bmatrix}
=\sum_K(-1)^{k_2+k_3+K}(2K+1)\begin{Bmatrix} J' & J & K\\ I & I &F \end{Bmatrix} \begin{Bmatrix} k_2 & k_3  & K\\ J & J' & J'' \end{Bmatrix} \begin{Bmatrix} k_2 & k_3  & K\\ I & I & I \end{Bmatrix}.
\end{equation*}
This increases the number of 6j-symbols in Eq.~\ref{eq:P} by one but reduces the number with an $F$-dependence to just two.  Applying the identity again reduces this to just one, which is exactly the factor required for Eq.~\ref{eq:hyperfineShift}. Treating Eq.~\ref{eq:G} in a similar way provides the desired expansion with $U_k^{(3)}(J)$ given by
\begin{multline}
\label{eq:3rdOrder1}
U_k^{(3)}(J) = \left(\begin{matrix}
I & k & I \\
-I & 0 & I
\end{matrix}\right)\left(\begin{matrix}
J & k & J \\
-J & 0 & J
\end{matrix}\right)\sum_{J'J''}\frac{(2k+1)}{(E_J-E_{J'})(E_J-E_{J''})}\sum_{k_1k_2k_3}\sum_{K}(-1)^{2K+\alpha}(2K+1)\\
\times\Bigg[
(-1)^{J-J''}
\left\{\begin{matrix}
k_1 & J & J' \\
J & K & k
\end{matrix}\right\}
\left\{\begin{matrix}
k_1 & I & I \\
I & K & k
\end{matrix}\right\}
\left\{\begin{matrix}
k_2 & J' & J'' \\
J & k_3 & K
\end{matrix}\right\}
\left\{\begin{matrix}
k_2 & k_3 & K \\
I & I & I
\end{matrix}\right\}
\mathcal{I}^{(k_1)}\mathcal{I}^{(k_2)}\mathcal{I}^{(k_3)}\mathcal{Q}^{(k_1)}_{JJ'}\mathcal{Q}^{(k_2)}_{J'J''}\mathcal{Q}^{(k_3)}_{J''J} \\
- \delta_{J'J''} 
\left\{\begin{matrix}
k_1 & J & J \\
J & K & k
\end{matrix}\right\}
\left\{\begin{matrix}
k_1 & I & I \\
I & K & k
\end{matrix}\right\}
\left\{\begin{matrix}
k_2 & J & J' \\
J & k_3 & K
\end{matrix}\right\}
\left\{\begin{matrix}
k_2 & k_3 & K \\
I & I & I
\end{matrix}\right\}\mathcal{I}^{(k_1)}\mathcal{I}^{(k_2)}\mathcal{I}^{(k_3)}\mathcal{Q}^{(k_1)}_{JJ}\mathcal{Q}^{(k_2)}_{JJ'}\mathcal{Q}^{(k_3)}_{J'J}\Bigg],
\end{multline}
where $\alpha=3J'+3J''+2J+k+k_1+k_2+k_3$.  The summation of the 6j-symbols over $K$ can be conveniently written in terms of a 12j-symbol of the second kind \cite[pg 367]{varshalovich1988quantum} to give
\begin{multline}
U_k^{(3)}(J) = \begin{pmatrix}
I & k & I \\
-I & 0 & I
\end{pmatrix}\begin{pmatrix}
J & k & J \\
-J & 0 & J
\end{pmatrix}\sum_{\substack{J'J''\\k_1k_2k_3}}\frac{(-1)^\alpha (2k+1)}{(E_J-E_{J'})(E_J-E_{J''})} \\
\times \left[\begin{Bmatrix} - & J & k_3 & J''\\ I & - & I & k_2\\k_1 & J & -  & J'\\I & k & I &- \end{Bmatrix}\mathcal{Q}^{(k_1)}_{JJ'}\mathcal{Q}^{(k_2)}_{J'J''}\mathcal{Q}^{(k_3)}_{J''J}
- \delta_{J'J''}(-1)^{J'-J}\begin{Bmatrix} - & J & k_3 & J'\\ I & - & I & k_2\\k_1 & J & -  & J\\I & k & I &- \end{Bmatrix}\mathcal{Q}^{(k_1)}_{JJ}\mathcal{Q}^{(k_2)}_{JJ'}\mathcal{Q}^{(k_3)}_{J'J}\right] \mathcal{I}^{(k_1)}\mathcal{I}^{(k_2)}\mathcal{I}^{(k_3)}.
\end{multline}
\end{widetext}
In this form, properties of the 12j-symbol make it immediately clear that $k\leq k_1+k_2+k_3$ for a non-zero contribution.

The validity of the form of the perturbation to any order can be established by induction. For $n^{th}$-order perturbation theory, the most general term has 6j-symbols appearing in the form
\begin{multline}
\begin{Bmatrix}F & J & I \\k_1 & I & J^{(1)}\end{Bmatrix}\\
\times\left(\begin{Bmatrix}F & J^{(n-1)} & I \\k_n & I & J\end{Bmatrix}\prod_{j=2}^{n-1}\begin{Bmatrix}F & J^{(j-1)} & I \\k_{j} & I & J^{(j)}\end{Bmatrix}\right),
\end{multline}
with all others as a special case, just as $\mathcal{G}$ is a special case of $\mathcal{P}$ in third-order perturbation.  If it is assumed the term in parentheses can be reduced to a product of 6j-symbols with only one having an F-dependence given by
\begin{equation}
\begin{Bmatrix}F & J^{(1)} & I \\K & I & J\end{Bmatrix},
\end{equation}
the \textit{Biedenharn-Elliott} identity can be used to produce same $F$-dependent factor as the first order result.  Moreover, the same identity can also be used to inductively infer the validity the assumption to any order.
\bibliography{Lu577BIB}
\bibliographystyle{unsrt}
\end{document}